\documentclass[aps,twocolumn,showpacs]{revtex4}
\usepackage{graphicx}
\usepackage{color}
\usepackage{natbib}

\newcommand{\bra}{\left\langle}
\newcommand{\ket}{\right\rangle}
\newcommand{\tbox}[1]{\mbox{\tiny #1}}

\begin{document}

\title{Topological versus spectral properties of random geometric graphs}

\author{R. Aguilar-S\'anchez,$^1$ J. A. M\'endez-Berm\'udez$^{2,3}$, Francisco A. Rodrigues$^2$, and Jos\'e M. Sigarreta$^4$}

\affiliation{
$^1$Facultad de Ciencias Qu\'imicas, Benem\'erita Universidad Aut\'onoma de Puebla,
Puebla 72570, Mexico \\
$^2$Departamento de Matem\'{a}tica Aplicada e Estat\'{i}stica, Instituto de Ci\^{e}ncias 
Matem\'{a}ticas e de Computa\c{c}\~{a}o, Universidade de S\~{a}o Paulo - Campus de S\~{a}o Carlos, 
Caixa Postal 668, 13560-970 S\~{a}o Carlos, SP, Brazil \\
$^3$Instituto de F\'{\i}sica, Benem\'erita Universidad Aut\'onoma de Puebla,
Apartado Postal J-48, Puebla 72570, Mexico \\
$^4$Facultad de Matem\'aticas, Universidad Aut\'onoma de Guerrero,
Carlos E. Adame No.54 Col. Garita, Acapulco Gro. 39650, Mexico}


\begin{abstract}
In this work we perform a detailed statistical analysis of topological and spectral properties of 
random geometric graphs (RGGs); a graph model used to study the structure and dynamics 
of complex systems embedded in a two dimensional space. 
RGGs, $G(n,\ell)$, consist of $n$ vertices uniformly and independently 
distributed on the unit square, where two vertices are connected by an edge if their Euclidian 
distance is less or equal than the connection radius $\ell \in [0,\sqrt{2}]$.
To evaluate the topological properties of RGGs we chose two well-known topological indices,
the Randi\'c index $R(G)$ and the harmonic index $H(G)$. While we characterize the spectral 
and eigenvector properties of the corresponding randomly-weighted adjacency matrices by the 
use of random matrix theory measures: the ratio between consecutive eigenvalue 
spacings, the inverse participation ratios and the information or Shannon entropies $S(G)$.
First, we review the scaling properties of the averaged measures, topological and spectral, on
RGGs. 
Then we show that:
(i) the averaged--scaled indices, $\bra R(G) \ket$ and $\bra H(G) \ket$, are highly 
correlated with the average number of non-isolated vertices $\bra V_\times(G) \ket$; and
(ii) surprisingly, the averaged--scaled Shannon entropy $\bra S(G) \ket$ is also highly 
correlated with $\bra V_\times(G) \ket$.
Therefore, we suggest that very reliable predictions of eigenvector properties of RGGs
could be made by computing topological indices. 
\end{abstract}
\maketitle

\date{\today}



\section{Introduction}

Graphs have become an essential tool in nowadays complex systems modeling since they 
account for the underlying structure of real-world as well as synthetic systems of current interest.
Depending on their building rules, graphs can be classified as regular or random:
While regular graphs are constructed deterministically, random graphs follow probabilistic 
construction rules. There are many well-known models of random graphs. Among them, 
probably the most popular are: Erd\"os-R\'enyi random graphs, the scale-free network model 
of Barab\'asi and Albert and the small-world networks of Watts and Strogatz. In these
random graph models the embedding dimension is not a relevant parameter, however,
in some applications it is necessary to preserve the spatial component~\cite{B11} of the complex 
system under study. In such a case, Random Geometric Graphs (RGGs)~\cite{DC02,P03} have 
been used to study the structure and dynamics of spatially embedded complex systems.

In its original version, RGGs~\cite{G61} (named as random plane networks) consider uniformly 
and independently distributed vertices on the plane, where vertices are linked if they are closer 
than a fixed distance. Later, interesting variations of the original RGGs have been studied.
Among them we can mention {\it soft} RGGs (see e.g.~\cite{P16,DG16} and references therein), 
where the connection rule between vertices becomes a probabilistic function, 
and {\it random rectangular graphs}~\cite{ES15}, which were introduced as a generalization of 
RGGs to allow for the flexibility of the embedding space. Moreover, RGGs have found 
applications in the study of social~\cite{WPR06} and neural networks~\cite{SF92}, 
synchronization phenomena~\cite{DGMN09}, wireless ad-hoc 
communications~\cite{Nekovee07,HAB09,MA13,DOP18} and disease dynamics~\cite{TG07,BAK08}.
Many properties of RGGs and its variations are already known. Topological properties such as 
the average degree, the degree distribution, the average path length and the clustering coefficient 
were reported in~\cite{ES15}.
The effects of geometrical boundaries on RGGs were aso discussed in~\cite{CDG12}.
In addition, some spectral properties of RGGs have already been reported, both 
theoretically~\cite{BEJ07} and numerically~\cite{BEJ07,NGB14,DGK16,AMGM18}.

In this work, within a random-matrix-theory (RMT) approach, we perform a statistical study of 
both topological and spectral properties of RRGs.
In particular, to evaluate the topological properties of RGGs we chose two well-known topological 
indices, the Randi\'c index and the harmonic index. While we characterize the spectral 
and eigenvector properties of the corresponding randomly-weighted adjacency matrices by the 
use of RMT measures: the ratio between consecutive eigenvalue 
spacings, the inverse participation ratios and the information or Shannon entropies.
Moreover, we report an unexpected high correlation between topological indices and 
the Shannon entropies of eigenvectors; thus, suggesting that very reliable predictions of 
eigenvector properties of RGGs could be made by just computing topological indices.

\subsection{The randomly-weighted graph model}

We consider RGGs $G(n,\ell)$ consisting of $n$ vertices uniformly and independently distributed 
on the unit square, where two vertices are connected by an edge if their Euclidian distance is less 
or equal than the connection radius $\ell \in [0,\sqrt{2}]$.
Here we will follow a recently introduced approach under which the adjacency matrices of random 
graphs are represented by RMT ensembles; see the application of this approach on 
Erd\"os-R\'{e}nyi graphs~\cite{MAM15}, random rectangular graphs~\cite{AMGM18}, 
$\beta$-skeleton graphs~\cite{AME19}, multiplex and multilayer networks~\cite{MFMR17}, and 
bipartite graphs~\cite{MAMPS19}.
Accordingly, we define the elements of the adjacency matrix $\mathbf{A}$ of the RGGs, 
$G(n,\ell)$, as
\begin{equation}
A_{ij}=\left\{
\begin{array}{cl}
\sqrt{2} \epsilon_{ii} \ & \mbox{for $i=j$}, \\
\epsilon_{ij} & \mbox{if there is an edge between vertices $i$ and $j$},\\
0 \ & \mbox{otherwise}.
\end{array}
\right.
\label{Aij}
\end{equation}
We choose $\epsilon_{ij}$ as statistically-independent random variables drawn from a normal 
distribution with zero mean and unity variance. Also, $\epsilon_{ij}=\epsilon_{ji}$, since $G$ is assumed 
as undirected. 
According to this definition, diagonal random matrices are obtained for $\ell=0$ (Poisson ensemble (PE), 
in RMT terms), whereas the Gaussian Orthogonal Ensemble (GOE) (i.e.~full real and symmetric random 
matrices) is recovered when $\ell=\sqrt{2}$. Therefore, a transition from the PE to the GOE can be 
observed by increasing $\ell$ from zero to $\sqrt{2}$, for any given graph size $n$.

We stress that the random weights we impose to the adjacency matrix $\mathbf{A}$, as defined 
in Eq.~(\ref{Aij}), do not play any role in the computation of vertex-degree-based indices, however they 
help us obtaining non-null adjacency matrices (that we can still diagonalize) for graphs with isolated 
vertices; so we can safely explore spectral and eigenvector properties in the limit $\ell\to 0$. 
Moreover, including random weights to the standard RGG model allows us to retrieve well known 
random matrices in the appropriate limits in order to use RMT results as a reference.

\section{Measures}

\subsection{Topological measures}

Topological indices based on end-vertex degrees of edges have been used for more than 40 years 
and some of them are recognized tools in chemical research. Probably, the best known among such 
descriptors are the Randi\'c connectivity index and the Zagreb indices.

Given a simple connected graph $G=(V(G),E(G))$ with the vertex set $V(G)$ and the edge set $E(G)$,
the \emph{Randi\'c connectivity index} was defined in~\cite{R} as
\begin{equation}
\label{R}
R(G) = \sum_{uv\in E(G)} \frac1{\sqrt{d_u d_v}} \ ,
\end{equation}
where $uv$ denotes the edge of the graph $G$ connecting the vertices $u$ and $v$, and $d_u$ is 
the degree of the vertex $u$. There are hundreds of papers and a couple of books dealing 
with $R(G)$ (see, e.g.,~\cite{GF,LG,LS} and the references therein).
In addition to the multiple applications of the Randi\'c index in physical chemistry, this index has found several 
applications in other research areas and topics, such as information theory \cite{GFK18}, network similarity 
\cite{NJ03}, protein alignment \cite{R15}, network heterogeneity \cite{E10}, and network robustness \cite{MMR17}. 
Moreover, in \cite{CDE15} the concept of graph entropy for weighted graphs was introduced, especially the 
Randi\'c weights. 

Other index that has attracted great interest in the last years is the \emph{harmonic index}. It is 
given by~\cite{Faj}
\begin{equation}
\label{H}
H(G) = \sum_{uv\in E(G)}\frac{2}{d_u + d_v} \ .
\end{equation}
See examples of recent studies of $H(G)$ in Refs.~\cite{MMRS21,DBAV,RS4,WTD,Z,ZX}.

From definitions~(\ref{R}-\ref{H}), when $\ell=0$ (i.e.~when all vertices in the RGG are 
isolated) we have $R(G)=0$ and $H(G)=0$, while for $\ell=\sqrt{2}$ (i.e.~when the RGG graph is 
complete), $R(G)=n/2$ and $H(G)=n/2$. That is, we expect to observe a transition from 0 to
$n/2$ for the quantities $\left< R(G) \right>$ and $\left< H(G) \right>$ when increasing $\ell$ from
0 to $\sqrt{2}$.

We want to stress that the statistical study of topological indices we perform here is justified by 
the random nature of the RGGs.
Since a given parameter pair $(n,\ell)$ represents an infinite-size ensemble of random graphs, 
the computation of a topological index on a single graph is irrelevant. In contrast, the computation
of a given topological index on a large ensemble of random graphs, all characterized by the same 
parameter pair $(n,\ell)$, may provide useful {\it average} information about the full ensemble.
This {\it statistical} approach, well known in RMT studies, is not widespread in 
studies of topological indices, mainly because topological indices are not commonly applied to 
random graphs; for recent exceptions see~\cite{MMRS20,MMRS21}.

\subsection{Spectral measures}

In this work we characterize the spectral and eigenvector properties of the randomly-weighted 
adjacency matrices of Eq.~(\ref{Aij}) by the use of well-known RMT measures: the ratio between 
consecutive eigenvalue spacings $r$, the inverse participation ratios $\mbox{IPR}$ and the 
information or Shannon entropies $S$.

On the one hand, given the ordered spectra $\{ \lambda_i \}$, the ratio $r_i$ is given by~\cite{ABG13}
\begin{equation}
\label{r}
r_i = \frac{\min(\lambda_{i+1}- \lambda_i,\lambda_{i}- \lambda_{i-1})}{\max(\lambda_{i+1}- \lambda_i,\lambda_{i}- \lambda_{i-1})} \ ,
\end{equation}
$i=2,\ldots,n-1$.
It is known that $\left< r \right>_{\tbox{PE}}\approx 0.3863$, while 
$\left< r \right>_{\tbox{GOE}}\approx 0.5359$~\cite{ABG13}. 
Here and below $\left< \cdot \right>$ denotes the average 
over all eigenvalues/eigenvectors of an ensemble of adjacency matrices $\mathbf{A}$. 
On the other hand, for the normalized eigenvector $\Psi^i$, i.e. $\sum_{j=1}^n | \Psi^i_j |^2 =1$, we have
\begin{equation}
\label{IPR}
\mbox{IPR}_i = \left[ \sum_{j=1}^n \left| \Psi^i_j \right|^4 \right]^{-1}
\end{equation}
and
\begin{equation}
\label{S}
S_i = -\sum_{j=1}^n \left| \Psi^i_j \right|^2 \ln \left| \Psi^i_j \right| ^2 \ .
\end{equation}

It is fair to mention that both $\mbox{IPR}$ and $S$, which quantify the extension of eigenvectors
in a given basis, have been widely used to study the localization characteristics of the eigenvectors 
of random graphs and network models. 
Among the vast amount of studies available in the literature, as examples of recent studies were the
$\mbox{IPR}$ and $S$ were applied on graphs studies, we can mention that:
(i) the $\mbox{IPR}$ facilitated the introduction of the concept of layer localization in multilayer random 
networks, this new concept was shown to have relevant implications in the dynamics of desease 
contagion in multiplex systems~\cite{ACPRM17,AMRM20}; also 
(ii) the $\mbox{IPR}$ allowed to demonstrate that the eigenvectors of random networks with 
power-law decaying bond strengths are multifractal objects~\cite{VMF19}; while 
(iii) $S$ was used to define universal parameters able to scale the eigenvector properties of 
multiplex and multilayer networks~\cite{MFMR17} and bipartite graphs~\cite{MAMPS19}.
In contrast, $r$ has been scarcely used in graph studies; for a recent exception see 
Ref.~\cite{TFM20}, were $P(r)$ served to characterize the percolation transition 
in weighted Erd\"os-R\'{e}nyi graphs.
We believe that the lack of use of $r$ in graph studies is mainly due to the fact that the introduction
of $r$ is relatively recent. In fact, most studies of spectral properties of random graphs, from a RMT 
point of view, are based on the nearest-neighbor energy level spacing distribution $P(s)$, see 
e.g.~\cite{MAM15} and the references therein.
However, we emphasize that we prefer to use $\left< r \right>$, instead of $P(s)$, because 
the calculation of the ratios $r_i\equiv \min(s_i, s_{i+1})/\max(s_i, s_{i+1})$ (with 
$s_i=(\lambda_{i+1}-\lambda_i)/\Delta$, $\Delta$ being the mean eigenvalue spacing) do not 
require the spectrum unfolding, a task that may become cumbersome. Moreover, the spectrum 
unfolding fixes $\left< s \right>=1$ and forbids the use of $\left< s \right>$ as a complexity indicator;
a restriction not applicable to $\left< r \right>$.
  
With definitions (\ref{IPR}-\ref{S}), when $\ell=0$, since the eigenvectors of $\mathbf{A}$ have only 
one nonvanishing component with magnitude equal to one, $\mbox{IPR}_i = 1$ and $S_i = 0$, so that 
$\left< \mbox{IPR}(G) \right>= \mbox{IPR}_{\tbox{PE}}=1$ and $\left< S(G) \right>= S_{\tbox{PE}}=0$, 
respectively. 
In the opposite limit, $\ell=\sqrt{2}$, the fully chaotic eigenvectors extend over the $n$ available vertices 
of $G$, so $\left< \mbox{IPR}(G) \right> = \mbox{IPR}_{\tbox{GOE}}$ and 
$\left< S(G) \right> = S_{\tbox{GOE}}$. Here, $\mbox{IPR}_{\tbox{GOE}}\approx n/3$ and 
$S_{\tbox{GOE}}\approx \ln (n/2.07)$ correspond to random eigenvectors with Gaussian distributed 
amplitudes; i.e.~eigenvectors of the GOE.
Thus, we expect to observe a transition from the PE to the GOE limits for $\left< r \right>$, 
$\left< \mbox{IPR}(G) \right>$ and $\left< S \right>$ when increasing $\ell$ from 0 to $\sqrt{2}$.

\section{Scaling of topological and spectral measures}

\subsection{Topological measures on RGGs}

Recall that increasing $\ell$ from 0 to $\sqrt{2}$ drives the RGGs from isolated vertices 
to complete graphs. Indeed, the set of non-isolated vertices of the RGGs, $V_\times(G)\subseteq V(G)$, 
are those that contribute to the sums in Eqs.~(\ref{R}-\ref{H}). Clearly, $V_\times(G)=0$
for $\ell=0$ while $V_\times(G)=n$ for $\ell=\sqrt{2}$, however the quantity $V_\times(G)$ as 
a function of $\ell$ is unknown (to the best of our knowledge) for RGGs. Therefore, together 
with $R(G)$ and $H(G)$, below we also compute $V_\times(G)$.

\begin{figure*}[t!]
\begin{center} 
\includegraphics[width=0.7\textwidth]{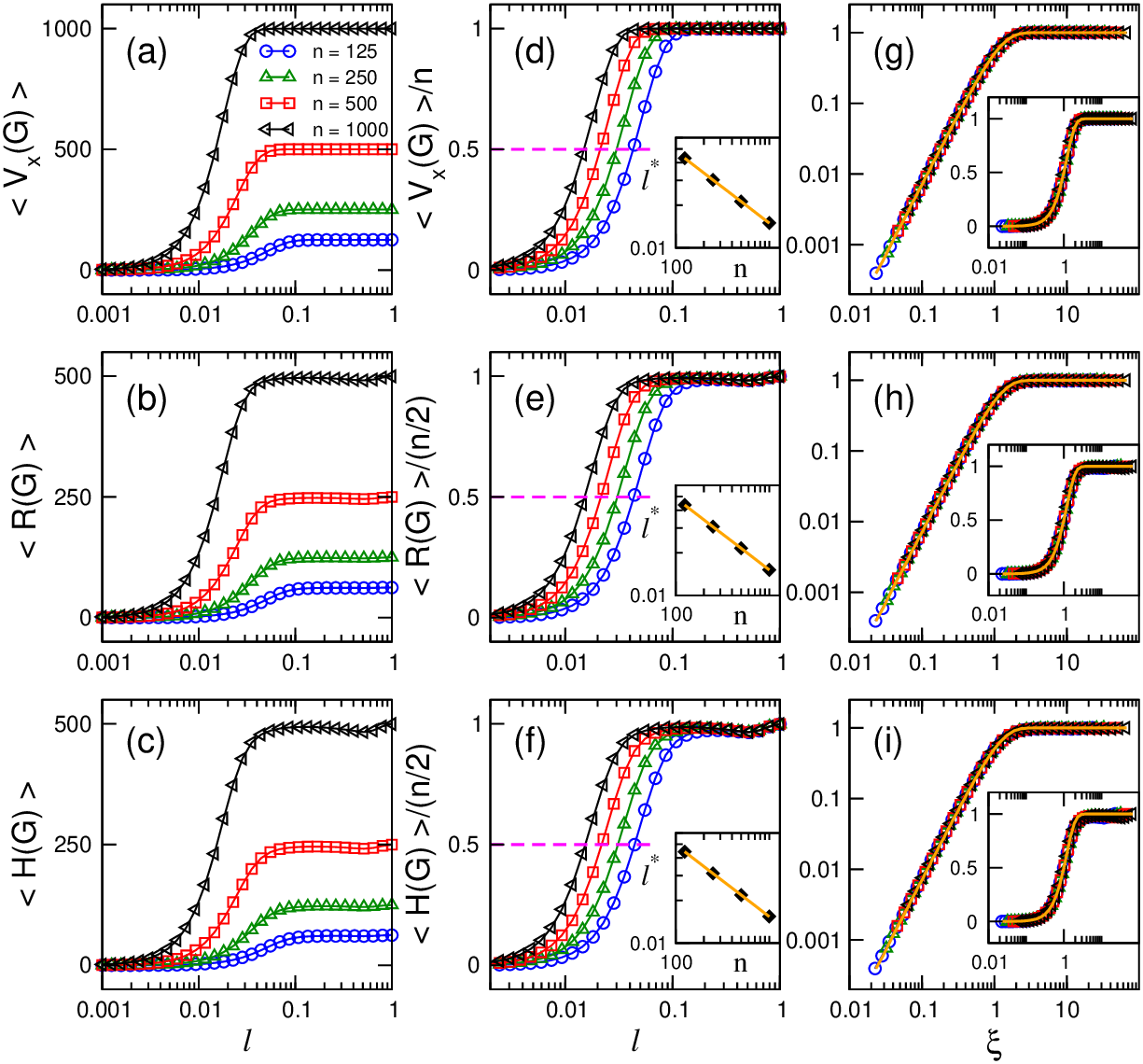}
\caption{
(a) Average number of non-isolated vertices $\bra V_\times(G) \ket$,
(b) average Randi\'c index $\left< R(G) \right>$, and
(c) average harmonic index $\left< H(G) \right>$
as a function of the connection radius $\ell$ of random geometric graphs of four different sizes $n$.
Normalized average indices $\bra V_\times(G) \ket/n$, $\left< R(G) \right>/(n/2)$ and 
$\left< H(G) \right>/(n/2)$ as a function of (d-f) $\ell$ and (g-i) $\xi$.
The horizontal dashed line in panels (d-f) marks $\left< \overline{X}(G) \right>=0.5$.
The insets in panels (d-f) show $\ell^*$ vs.~$n$; the orange lines represent the best fittings of the data
with Eq.~(\ref{scaling}), with fitting parameters reported in Table~\ref{Table1}.
The insets in panels (g-i) show the same curves of the main panels but in semi-log scale. 
The orange line, shown for comparison purposes in the main panels (g-i) and the insets, is the 
curve $\bra V_\times(G) \ket/n$ vs.~$\xi$.
}
\label{Fig01}
\end{center}
\end{figure*}

Now, in Figs.~\ref{Fig01}(a-c) we present $\left< V_\times(G) \right>$, $\left< R(G) \right>$, and 
$\left< H(G) \right>$ as a function of the connection radius $\ell$ for RGGs of size $n$.
All averages are computed over ensembles of $10^7/n$ RGGs characterized by the parameter 
pair $(n,\ell)$. Each panel displays four curves corresponding 
to different graph sizes: 125, 250, 500 and 1000.
Note that all curves $\left< X(G) \right>$ vs.~$\ell$ in Figs.~\ref{Fig01}(a-c) display the transition 
(in fact, a smooth transition in semi-log scale) from isolated vertices to complete graphs when 
increasing $\ell$. In this Section, $X$ stands for $V_\times$, $R$ and $H$.

\begin{table}[b!]
\begin{center}
\caption{Values of the parameters $\cal{C}$ and $\gamma$ obtained from the fittings of Eq.~(\ref{scaling})
to the data $\ell^*$ vs.~$n$ of the insets in panels (d-f) of Figs.~\ref{Fig01} and~\ref{Fig03}.}
\vspace{0.25cm}
\begin{tabular}{ l  l  l  l  l } 
\hline
measure & \qquad \qquad & $\cal{C}$ & \qquad \qquad & $\gamma$  \\
\hline \hline
$\bra V_\times(G) \ket/n$                  & & 0.49794  & &  0.50751 \\ \hline
$\bra R(G) \ket/(n/2)$                        & & 0.50519  & &  0.50756 \\ \hline
$\bra H(G) \ket/(n/2)$                        & & 0.51253  & &  0.50764 \\ \hline
$\bra \overline{r}(G) \ket$                  & & 1.2883    & &  0.44193 \\ \hline
$\bra \overline{IPR}(G) \ket$             & & 1.1844    & & 0.36272 \\ \hline
$\left< S(G) \right>/S_{\tbox{GOE}}$ & & 1.0463    & & 0.44138 \\ \hline
\end{tabular}
\label{Table1}
\end{center}
\end{table}

Given the similar functional form of the curves $\left< X(G) \right>$ vs.~$\ell$ for different graph
sizes $n$, as reported in Figs.~\ref{Fig01}(a-c), it seems that they could be effectively scaled. 
That is, one should be able to find a scaling parameter $\xi\equiv\xi(n,\ell)$ such that the curves 
$\left< \overline{X}(G) \right>$ vs.~$\xi$ are invariant, where $\overline{X}$ is the properly 
normalized measure $X$. We choose to normalize the topological indices $X$ to their maximum 
values: $\max[V_\times(G)]=n$, $\max[R(G)]=n/2$ and $\max[H(G)]=n/2$. Then, in 
Figs.~\ref{Fig01}(d-f) we present again $\bra X(G) \ket$ but now normalized to $\max[X(G)]$.
From these figures we can clearly see that when changing $n$ the curves 
$\bra \overline{X}(G) \ket =\bra X(G) \ket/\max[X(G)]$ keep their functional form but they suffer 
a displacement on the 
$\ell$-axis. Therefore, in order to search for the scaling parameter $\xi\equiv \xi(n,\ell)$ 
we first establish a quantity to characterize the position of the curves $\bra \overline{X}(G) \ket$ 
on the $\ell$-axis. Since all curves $\bra \overline{X}(G) \ket$ vs.~$\ell$ transit from zero 
(isolated vertices) to one (complete graphs) when $\ell$ increases from zero to $\sqrt{2}$, we 
choose the value of $\ell$ for which $\bra \overline{X}(G) \ket \approx 0.5$; see the horizontal 
dashed lines in Figs.~\ref{Fig01}(d-f). We label the value of $\ell$ at half of the transition as 
$\ell^*$.

In the insets of Figs.~\ref{Fig01}(d-f) we present the values of $\ell^*$, extracted from the curves 
of the main panels, versus $n$. Indeed, the linear trend of the data $\ell^*$ vs.~$n$ (in log-log 
scale) suggests the power-law behavior
\begin{equation}
\label{scaling}
\ell^* = \mathcal{C}n^{-\gamma} .
\end{equation}
As shown in the insets of Figs.~\ref{Fig01}(d-f), Eq.~(\ref{scaling}) provides excellent fittings to 
the data; see the full-orange lines. From the fitted parameters, reported in Table~\ref{Table1}, we 
can conclude that $\gamma\approx 1/2$ for the three indices: $V_\times(G)$, $R(G)$ and $H(G)$.

\begin{figure*}[t!]
\begin{center} 
\includegraphics[width=0.7\textwidth]{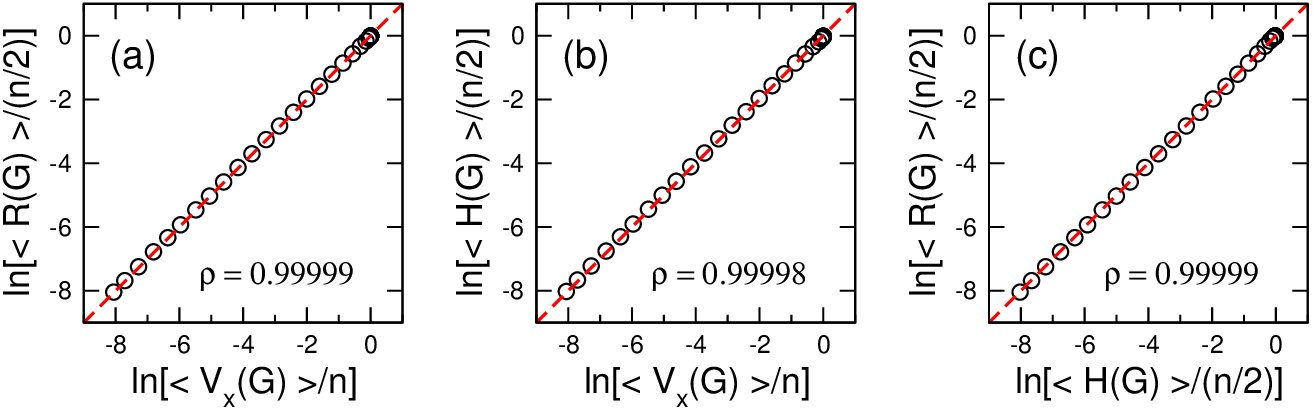}
\caption{
Scatter plots of the log of the normalized indices reported in Fig.~\ref{Fig01}.
The red-dashed lines, shown as a reference, are the identity function.
The Pearson correlation coefficients are reported in the corresponding panels.
}
\label{Fig02}
\end{center}
\end{figure*}

Finally, we define the scaling parameter $\xi$ as the ratio between $\ell$ and $\ell^*$, so we get  
\begin{equation}
\label{xi}
\xi \equiv \frac{\ell}{\ell^*} = \frac{\ell}{{\cal C}n^{-\gamma}} = \frac{n^\gamma \ell}{\cal{C}} \ .
\end{equation} 
Therefore, by plotting again the curves of Figs.~\ref{Fig01}(d-f) now as a function of $\xi$
we observe that curves for different graph sizes $n$ collapse on top of {\it universal curves}; see 
Figs.~\ref{Fig01}(g-i) where we present $\bra X(G) \ket/\max[X(G)]$ vs.~$\xi$ in log-log scale
(main panels) as well as in semi-log scale (insets).
Also note that the three indices are characterized by very similar universal curves.
Indeed, in Figs.~\ref{Fig01}(h) and~\ref{Fig01}(i) we plot the curve $\bra V_\times \ket/n$ vs.~$\xi$ 
on top of the data for $\bra R(G) \ket$ and $\bra H(G) \ket$, respectively, 
and observe a remarkably good coincidence anticipating a very high correlation among these
topological indices.
Moreover, in Fig.~\ref{Fig02} we present scatter plots between the (normalized and scaled) 
indices reported in 
Fig.~\ref{Fig01}, where the high correlation between them is evident. To quantify the correlation
among these indices, in the panels of Fig.~\ref{Fig02} we report the corresponding Pearson's 
correlation coefficient $\rho$, which turns out to be approximately equal to one in all cases.
We note that we compute $\rho$ for the log of the variables to avoid an overestimation caused 
by the data with $\bra \overline{X}(G) \ket\sim 1$.

Figure~\ref{Fig02} provides clear evidence of the relevance of the non-isolated vertices
$V_\times(G)$ in the computation of topological indices, particularly in the computation of 
$R(G)$ and $H(G)$. That is, once knowing the value of the parameter $\mathcal{C}$, one 
could compute $\bra V_\times(G) \ket$ and make quite precise predictions of 
$\bra R(G) \ket$ or $\bra H(G) \ket$ in RGGs.

\subsection{Spectral measures on RGGs}
\label{spectralscaling}

Now we perform a scaling study of spectral and eigenvector measures of RGGs 
equivalent to the one made in the previous Section for topological indices.
We use exact numerical diagonalization to obtain the eigenvalues $\lambda_i$ and 
eigenvectors $\Psi^i$ ($i =1,\ldots,n$) of large ensembles of adjacency matrices given 
by Eq.~(\ref{Aij}) (characterized by $n$ and $\ell$) and compute the average values of
the measures of Eqs.~(\ref{r}-\ref{S}).

\begin{figure*}[t!]
\begin{center} 
\includegraphics[width=0.7\textwidth]{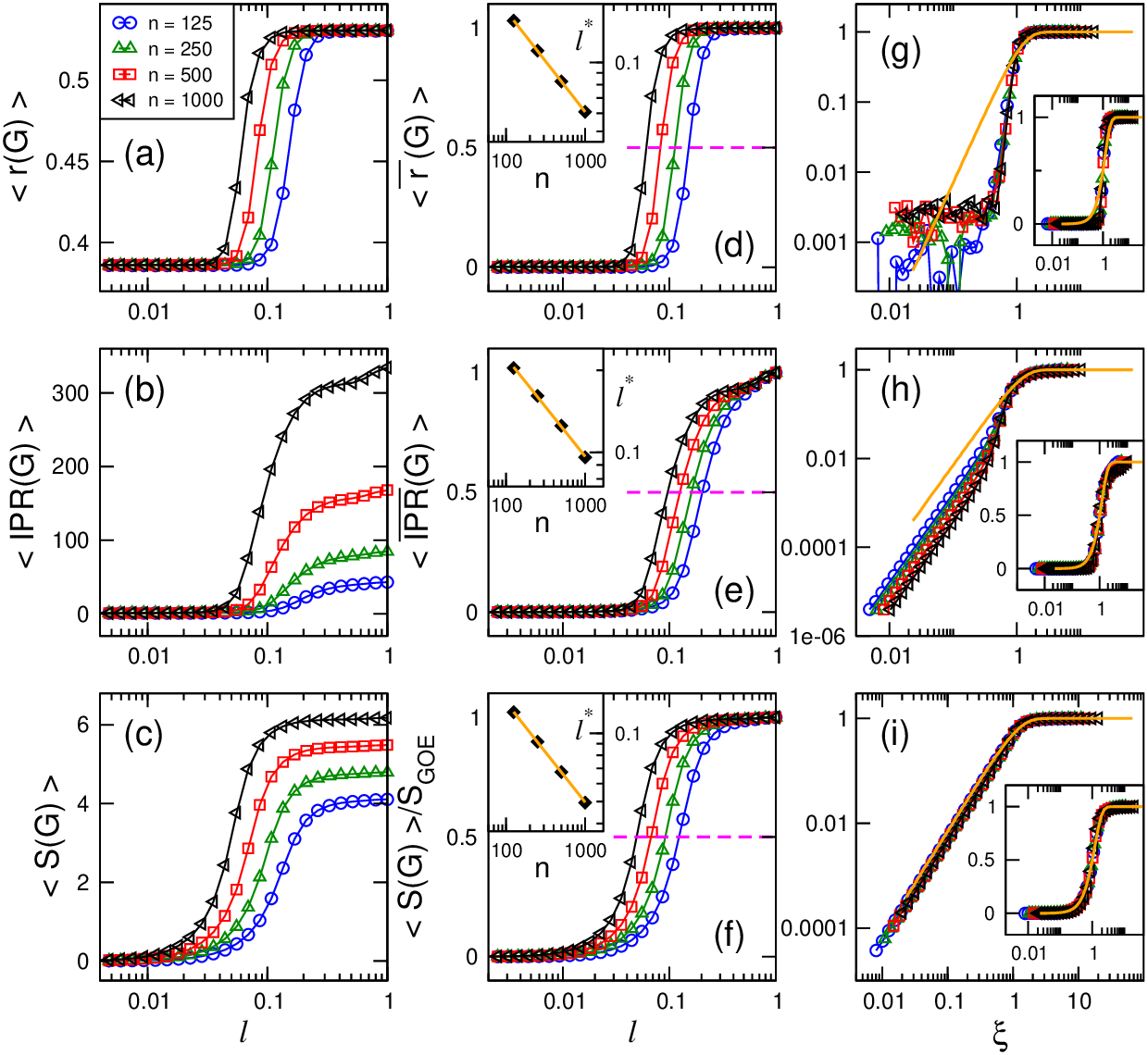}
\caption{
(a) Average ratio between consecutive eigenvalue spacings $\bra r(G) \ket$,
(b) average inverse participation ratio $\left< \mbox{IPR}(G) \right>$, and
(c) average Shannon entropy $\left< S(G) \right>$
as a function of the connection radius $\ell$ of random geometric graphs of four different sizes $n$.
Normalized average measures $\bra \overline{r}(G) \ket$, $\bra \overline{\mbox{IPR}}(G) \ket$ and 
$\left< S(G) \right>/S_{\tbox{GOE}}$ as a function of (d-f) $\ell$ and (g-i) $\xi$.
The insets in panels (d-f) show $\ell^*$ vs.~$n$; the orange lines represent the best fittings of the data
with Eq.~(\ref{scaling}), with fitting parameters reported in Table~\ref{Table1}.
The horizontal dashed line in panels (d-f) marks $\left< \overline{X}(G) \right>=0.5$.
The insets in panels (g-i) show the same curves of the main panels but in semi-log scale. 
The orange line, shown for comparison purposes in the main panels (g-i) and the insets, is the 
curve $\bra V_\times(G) \ket/n$ vs.~$\xi$.
}
\label{Fig03}
\end{center}
\end{figure*}

In Figs.~\ref{Fig03}(a-c) we present $\bra r(G) \ket$, $\left< \mbox{IPR}(G) \right>$ and
$\left< S(G) \right>$ as a function of the connection radius $\ell$ for RGGs of size $n$.
All averages are computed over all eigenvalues or eigenvectors of ensembles of $10^7/n$ RGGs.
Note that all curves $\left< X(G) \right>$ vs.~$\ell$ in Figs.~\ref{Fig03}(a-c) display the transition 
from the PE regime (isolated vertices; i.e~diagonal random matrices) to the GOE regime (complete 
graphs; i.e~full random matrices) for increasing $\ell$. In this Section $X$ stands for 
$r$, $\mbox{IPR}$ and $S$.

To perform the scaling analysis of the curves $\left< X \right>$ vs.~$\ell$ we first normalize them.
However, since $\left< r(G) \right>\to\mbox{const.}\ne 0$ and 
$\left< \mbox{IPR}(G) \right>\to\mbox{const.}\ne 0$ for $\ell\to 0$, we conveniently define the
corresponding normalized averages as 
$$
\left< \overline{r}(G) \right> \equiv \frac{\left< r(G) \right>-\left< r \right>_{\tbox{PE}}}{\left< r \right>_{\tbox{GOE}}-\left< r \right>_{\tbox{PE}}} 
$$
and
$$
\left< \overline{\mbox{IPR}}(G) \right> \equiv \frac{\left< \mbox{IPR}(G) \right>-\mbox{IPR}_{\tbox{PE}}}{\mbox{IPR}_{\tbox{GOE}}-\mbox{IPR}_{\tbox{PE}}} \ .
$$
While we normalize $\left< S(G) \right>$ with $S_{\tbox{GOE}}$: 
$\left< \overline{S}(G) \right>=\left< S(G) \right>/S_{\tbox{GOE}}$. Then, in Figs.~\ref{Fig03}(d-f) 
we present the normalized measures $\bra \overline{X}(G) \ket$ as a function of $\ell$.
From these figures we can see that when increasing $n$, the curves $\bra \overline{X}(G) \ket$ keep 
their functional form but suffer a displacement to the left on the $\ell$-axis. Also, since all curves 
$\bra \overline{X}(G) \ket$ vs.~$\ell$ transit from zero (PE regime) to one (GOE regime) when $\ell$ 
increases from zero to $\sqrt{2}$, as in the previous Section we can use the
value of $\ell$ for which $\bra \overline{X}(G) \ket \approx 0.5$, $\ell^*$, to 
characterize the position of the curves $\bra \overline{X}(G) \ket$ vs.~$\ell$ on the $\ell$-axis. 
Notice that $\ell^*$ characterizes the PE--to--GOE transition of the RGGs.

\begin{figure*}[t!]
\begin{center} 
\includegraphics[width=0.7\textwidth]{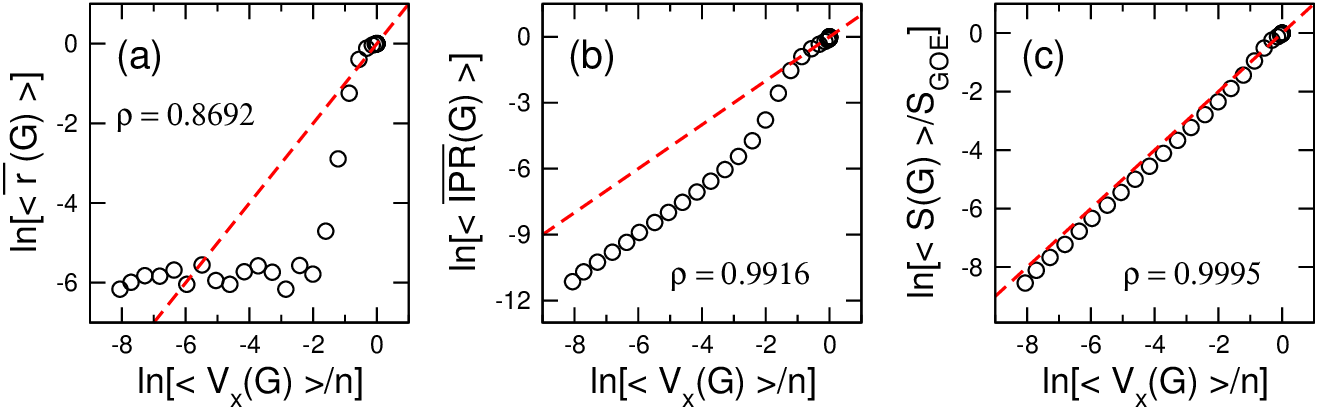}
\caption{
Scatter plots of the log of the normalized spectral measures reported in Fig.~\ref{Fig03} versus the 
log of the normalized average number of non-isolated vertices $\bra V_\times(G) \ket/n$.
The red-dashed lines, shown as a reference, are the identity function.
The Pearson correlation coefficients are reported in the corresponding panels.
}
\label{Fig04}
\end{center}
\end{figure*}

In the insets of Figs.~\ref{Fig03}(d-f) we plot the values of $\ell^*$, extracted from the curves 
of the main panels, versus $n$. Again as for the topological indices, the linear trend of the data 
$\ell^*$ vs.~$n$ (in log-log scale) suggests the power-law behavior of Eq.~(\ref{scaling}).
Therefore we use Eq.~(\ref{scaling}) to fit the data in the insets of Figs.~\ref{Fig03}(d-f), obtaining
excellent fittings. The fitted parameters are reported in Table~\ref{Table1}. Here, in contrast to
the topological indices studied in the previous Section, we do not find the same power law $\gamma$
for all spectral measures: i.e.~$\gamma\approx 0.44$ for $r(G)$ and $S(G)$, while 
$\gamma\approx 0.36$ for $\mbox{IPR}(G)$. Anyway, to scale the spectral measures on RGGs 
we use the scaling parameter $\xi$, as defined in Eq.~(\ref{xi}). 

Thus, in Figs.~\ref{Fig03}(g-i) we plot again the curves of Figs.~\ref{Fig03}(d-f) but now as 
a function of $\xi$. It is interesting to note that, even though we observe a perfect collapse of the
curves $\bra \overline{r}(G) \ket$ vs.~$\xi$ and $\bra S(G) \ket/S_{\tbox{GOE}}$ vs.~$\xi$
(except for large fluctuations in the curves of $\bra \overline{r}(G) \ket$ vs.~$\xi$ for small 
$n$ and small $\ell$), the average inverse participation ratio does not scale properly for $\ell<0.5$; 
of course this effect is only visible in the log-log scale (see the insets where the scaling looks 
reasonably good).  

Also, notice from Figs.~\ref{Fig03}(g) and~\ref{Fig03}(i) that the universal curves characterizing 
$\bra \overline{r}(G) \ket$ and $\bra S(G) \ket/S_{\tbox{GOE}}$ are clearly different.
Moreover, it results quite remarkable for us to observe that the universal curve of 
$\bra S(G) \ket/S_{\tbox{GOE}}$ looks similar to the universal curves of the topological indices
reported in the previous Section. Indeed, in Figs.~\ref{Fig03}(h-i) we plot the curve 
$\bra V_\times(G) \ket/n$ vs.~$\xi$ on top of the data for the spectral and eigenvector measures 
and observe a very 
good coincidence with the scaled Shannon entropy.
Also, in Fig.~\ref{Fig03} we present scatter plots between the normalized measures 
reported in Fig.~\ref{Fig03} and the scaled average number of non-isolated vertices. 
There the correlation between $S(G)$ and $V_\times(G)$ is evident, with a Pearson's 
correlation coefficient $\rho\approx 1$.

\section{Conclusions}
\label{Conclusions}

In this work we unveil an important link between topological and eigenvector properties of random 
geometric graphs (RGGs). Specifically, we numerically show that the Shannon entropies $S(G)$ of 
the eigenvectors of the randomly-weighted adjacency matrices of RGGs are highly correlated with 
the number of non-isolated graph vertices $V_\times(G)$ as well as with the Randi\'c index $R(G)$ 
and the harmonic index $H(G)$. Thus, we suggest that reliable predictions of eigenvector 
properties of RGGs could be made by computing topological indices. 
Clearly, an open question we plan to address in a future investigation is whether the correlation 
between topological indices and Shannon entropies also appears in other random network models.

Recently, it was demonstrated that the parameter that scales a large variety of average topological 
indices, including $R(G)$ and $H(G)$, on random graphs is the average degree~\cite{AHMS20};
which for RGGs is given by~\cite{ES15} $\bra k \ket=(n-1)(\pi \ell^2 - 8\ell^3/3 + \ell^4/2)$. Instead,
here we report $\xi \propto n^{1/2}\ell$ as the scaling parameter of $V_\times(G)$, $R(G)$ and $H(G)$. 
This apparent mismatch can be understood by realizing that not only $\bra k \ket$ but any function 
of it should scale end-vertex-degree based topological indices on random graphs; thus, since
$\xi \propto \bra k \ket^{1/2}$, for $\ell\ll 1$, our results agree with those in~\cite{AHMS20}.
It is interesting to highlight, however, that the scaling parameter of the spectral and eigenvector 
properties of RGGs can not be written in terms of $\bra k \ket$, as clearly shown in Table~\ref{Table1};
thus, validating the need of the scaling study developed in Sec.~\ref{spectralscaling}.

We hope that our work may motivate further analytical as well as numerical studies on
the applications of topological indices to random networks.


\section{Acknowledgments}

J.A.M.-B. acknowledges financial support from FAPESP (Grant No.~2019/ 06931-2), Brazil, 
CONACyT (Grant No.~2019-000009-01EXTV-00067) and PRODEP-SEP 
(Grant No.~511-6/2019.-11821), Mexico.  
F.A.R. thanks CNPq (Grant No.~309266/2019-0) for the financial support given to this research.


\end{document}